\documentclass[apjl]{emulateapj}
\usepackage{psfig,amsfonts,amsmath,graphicx,natbib,apjfonts,subfigure}
\def\ra#1#2#3{#1$^{\rm h}$#2$^{\rm m}$#3$^{\rm s}$}
\def\dec#1#2#3{$#1^\circ#2'#3''$}
\def\nod{\nodata}
\def\grb{GRB\,100117A}

\def\har{1}
\def\lei{2}
\def\war{3}
\def\stsci{4}
\def\berk{5}
\def\psu{6}

\shorttitle{The Afterglow and Host of Short GRB\,100117A}
\shortauthors{Fong et al.}

\begin{document}

\title{The Optical Afterglow and $z=0.92$ Early-Type Host Galaxy of
the Short GRB\,100117A}

\author{ W.~Fong\altaffilmark{\har}, E.~Berger\altaffilmark{\har},
R.~Chornock\altaffilmark{\har}, N.R.~Tanvir\altaffilmark{\lei},
A.J.~Levan\altaffilmark{\war}, A.S.~Fruchter\altaffilmark{\stsci},
J.F.~Graham\altaffilmark{\stsci}, A.~Cucchiara\altaffilmark{\berk},
D.B.~Fox\altaffilmark{\psu} }

\altaffiltext{\har}{Harvard-Smithsonian Center for Astrophysics, 60
Garden Street, Cambridge, MA 02138}

\altaffiltext{\lei}{Department of Physics and Astronomy, University of
Leicester, University Road, Leicester LE1 7RH, UK}

\altaffiltext{\war}{Department of Physics, University of Warwick,
Coventry, CV4 7AL, UK}

\altaffiltext{\stsci}{Space Telescope Science Institute, 3700 San
Martin Drive, Baltimore, MD 21218}

\altaffiltext{\berk}{Lawrence Berkeley National Laboratory, MS
50B-4026, 1 Cyclotron Road, Berkeley, CA, 94720}

\altaffiltext{\psu}{Department of Astronomy and Astrophysics, 525
Davey Laboratory, Pennsylvania State University, University Park, PA
16802}

\begin{abstract}
We present the discovery of the optical afterglow and early-type host
galaxy of the short-duration \grb.  The faint afterglow is
detected 8.3 hr after the burst with $r_{\rm AB}=25.46\pm 0.20$ mag.
Follow-up optical and near-IR observations uncover a coincident
compact red galaxy, identified as an early-type galaxy at a
photometric redshift of $z\approx 0.6-0.9$ ($2\sigma$) with a mass of
$\sim 3\times 10^{10}$ M$_\odot$, an age of $\sim 1$ Gyr, and a
luminosity of L$_{\rm B}\simeq 0.5$L$_{\rm *}$.  Spectroscopic
observations of the host reveal a notable break corresponding to the
Balmer/4000\AA\ break at $z\approx 0.9$, and stellar population
spectral evolution template fits indicate $z\approx 0.915$, which we
adopt as the redshift of the host, with stellar population ages of
$\sim 1-3$ Gyr.  From a possible weak detection of
[\ion{O}{2}]$\lambda 3727$ emission at $z=0.915$ we infer an upper
bound on the star formation rate of $\sim 0.1$ M$_\odot$ yr$^{-1}$,
leading to a specific star formation rate of $\lesssim 0.004$
Gyr$^{-1}$.  Thus, \grb\ is only the second short burst to date with a
secure early-type host (the other being GRB\,050724 at $z=0.257$) and
it has one of the highest short GRB redshifts.  The offset between the
host center and the burst position, $470\pm 310$ pc, is the smallest
to date.  Combined with the old stellar population age, this indicates
that the burst likely originated from a progenitor with no significant
kick velocity.  However, from the brightness of the optical afterglow
we infer a relatively low density of $n\approx 3\times 10^{-4}\,
\epsilon_{e,-1}^{-3}\epsilon_{B,-1}^{-1.75}$ cm$^{-3}$.  The
combination of an optically faint afterglow and host suggest that
previous such events may have been missed, thereby potentially biasing
the known short GRB host population against $z\gtrsim 1$ early-type
hosts.
\end{abstract}


\section{Introduction}
\label{sec:intro}

Progress in our understanding of short-duration gamma-ray bursts
(GRBs) and their progenitors relies on detailed studies of their
afterglows and host galaxy environments.  Of particular interest are
bursts with precise sub-arcsecond positions, which can provide
unambiguous host galaxy associations, redshifts, and burst properties
(energy, density).  Such localizations require the detection of
ultraviolet, optical, near-infrared, and/or radio afterglows; or
alternatively an X-ray detection with the {\it Chandra} X-ray
Observatory.  As of December 2010, only 20 short bursts have been
precisely localized in this manner.  Of these, 14 have clearly
identified hosts\footnotemark\footnotetext{These bursts are 050709:
\citep{ffp+05,hwf+05}; 050724: \citep{bpc+05}; 051221A:
\citep{sbk+06}; 060121 \citep{ltf+06,pcg+06}; 060313: \citep{rvp+06};
061006: \citep{amc+09}; 070707: \citep{pac+08}; 070714B:
\citep{gfl+09}; 070724: \citep{bcf+09}; 071227: \citep{amc+09};
080905: \citep{rwl+10}; 090426: \citep{aap+09,lbb+10}; 090510:
\citep{mkr+10}; and 100117A: this paper.} (10 with spectroscopic
redshifts), 5 do not have unambiguous host
associations\footnotemark\footnotetext{GRBs 061201: \citep{sap+07},
070809: \citep{gcn6739}, 080503: \citep{pmg+09}, 090305:
\citep{gcn8934,gcn8933}, and 090515: \citep{rot+10}.} \citep{ber10a},
and 1 has not been reported in the literature so
far\footnotemark\footnotetext{GRB 091109b: \citep{gcn10156}.}. For a
recent summary see \citet{ber10b}.

Only in a single case out of the 10 hosts with spectroscopic
identifications is the galaxy known to be early-type with no evidence
for on-going star formation activity (GRB\,050724: \citealt{bpc+05});
the remaining hosts are star forming galaxies, albeit at a level that
is on average significantly lower than in long GRB hosts
\citep{ber09}, particularly when accounting for their higher
luminosities and stellar masses \citep{ber09,lb10}.  The hosts with
measured redshifts span a range of $z\approx 0.1-1$ (e.g.,
\citealt{ber09}), with the exception of GRB\,090426 at $z=2.609$
\citep{aap+09,lbb+10}; in the three remaining cases the hosts are too
faint for a spectroscopic redshift determination, but are likely to be
located at $z\gtrsim 1$ \citep{bfp+07}.  At the same time, there is
tentative evidence for early-type hosts in the sample of short bursts
with optical positions and no coincident hosts based on chance
coincidence probabilities (GRBs 070809 and 090515; \citealt{ber10a}).

The host demographics and redshift distribution provide important
constraints on the nature of the progenitors.  For example, an
abundance of low redshifts and early-type hosts would point to a
population that is skewed to old ages, $\gtrsim {\rm few}$ Gyr
\citep{zr07}.  However, from the existing host galaxy demographics and
redshift distributions it appears that the progenitors span a broad
range of ages, $\sim 0.1-{\rm few}$ Gyr, and may indeed be
over-represented in late-type galaxies with intermediate-age
populations ($\sim 0.3$ Gyr; \citealt{bfp+07,lb10}).

Afterglow detections are also important for determining the GRB and
circumburst medium properties.  To date, however, little detailed
information about these properties has been drawn from the existing
(though sparse) broad-band afterglow detections (e.g.,
\citealt{bpc+05,pan06,rvp+06,sbk+06,ber10a}), mainly due to the
faintness of short GRB afterglows.  Early time optical observations
also probe possible emission from radioactive material synthesized and
ejected in a binary compact object merger, a so-called Li-Paczynski
mini-supernova \citep{lp98,mmd+10}.  No such emission has been
conclusively detected to date (e.g., \citealt{hsg+05,bpp+06,bcf+09}).

Here we report the discovery of the optical afterglow and host galaxy
of the short \grb.  From spectroscopy and optical/near-IR
imaging we find that the host is an early-type galaxy at $z=0.915$,
making it only the second unambiguous early-type host association for
a short GRB with a significantly higher redshift than the previous
event, GRB 050724 at $z=0.257$ \citep{bpc+05}.  The precise position
also allows us to measure the burst offset, while the optical flux
provides constraints on the circumburst density.  We present the
afterglow and host discovery in \S\ref{sec:obs}.  In \S\ref{sec:host}
we study the host redshift and stellar population properties, while in
\S\ref{sec:ag} we analyze the afterglow properties.  Finally, we draw
conclusions about the nature of this burst and implications for the
short GRB sample in \S\ref{sec:discussion}.

Throughout the paper we use the standard cosmological parameters,
$H_{0}=71$ km s$^{-1}$ Mpc$^{-1}$, $\Omega_{m}=0.27$, and
$\Omega_{\Lambda}=0.73$.

\section{Observations of GRB\,100117A}
\label{sec:obs}

\grb\ was detected by the {\it Swift} satellite on 2010 January 17.879
UT, and an X-ray counterpart was promptly localized by the on-board
X-ray Telescope (XRT) with a final positional accuracy of $2.4''$
radius \citep{gcnr269}.  No optical/UV source was detected by the
co-aligned UV/Optical Telescope (UVOT).  The burst duration is
$T_{90}=0.30\pm 0.05$ s, and its $15-150$ keV fluence is
$F_\gamma=(9.3\pm 1.3)\times 10^{-8}$ erg cm$^{-2}$.  The burst was
also detected by the {\it Fermi} Gamma-Ray Burst Monitor (GBM) with a
duration of about 0.4 s, an $8-1000$ keV fluence of $F_\gamma=(4.1\pm
0.5)\times 10^{-7}$ erg cm$^{-2}$, and a peak energy of
$E_p=287^{+74}_{-50}$ keV \citep{gcnr269}. These properties
clearly classify \grb\ as a short burst.

The X-ray light curve exhibits a complex behavior at early time, with
an initial flare lasting until about 200 s, followed by a steep
decline with $F_X\propto t^{-3.5\pm 0.2}$.  Subsequent data collected
at $\sim 5-690$ ks lead to an upper limit of $F_X\lesssim 2.5\times
10^{-14}$ erg cm$^{-2}$ s$^{-1}$ (unabsorbed;
\citealt{ebp+07,ebp+09}).  A fit to the X-ray spectrum indicates that
$F_X\propto\nu^{-0.70\pm 0.14}$ and $N_H=(1.2\pm 0.4)\times 10^{21}$
cm$^{-2}$, in excess of the expected Galactic column of $N_H=2.7\times
10^{20}$ cm$^{-2}$ \citep{gcnr269}. Using the counts to unabsorbed
flux conversion ($1\,{\rm cps}=5.2\times 10^{-11}$ erg cm$^{-2}$
s$^{-1}$) and the measured spectral index, we find a flux density
limit at $\gtrsim 5$ ks of $F_\nu(1\,{\rm keV})\lesssim 2.3\times
10^{-3}$ $\mu$Jy.

Ground-based follow-up optical observations of the XRT position were
first obtained with the Nordic Optical Telescope (NOT) 20.4 min
after the burst and led to the marginal detection of a source with
$R\approx 22.5$ mag \citep{gcn10337}.  Inspection of archival optical
images from the Canada-France-Hawaii MegaCam survey led to the
detection of four faint sources within the initial XRT error circle
with $R,I\approx 23-24.5$ mag, while $i$-band imaging at about 4.7 hr
revealed that the brightest of these sources had a comparable
brightness to the archival data and was likely the same source
detected by \citealt{gcn10337} \citep{gcn10339}.

\subsection{Afterglow Discovery}
\label{sec:disc}

We initiated $R$-band observations of \grb\ with the Inamori Magellan
Areal Camera and Spectrograph (IMACS) on the Magellan/Baade 6.5-m
telescope on 2010 January 18.04 UT (3.9 hr after the burst) and
detected the four sources noted by \citet{gcn10339} within the initial
XRT error circle.  We subsequently obtained two deeper epochs of
imaging in the $r$-band with the Gemini Multi-Object Spectrograph
(GMOS-N) on the Gemini-North 8-m telescope on 2010 January 18.21 and
19.22 UT (7.9 and 33.2 hr after the burst); see Table~\ref{tab:obs}.
The IMACS observations were reduced using standard procedures in IRAF,
while the GMOS-N observations were analyzed using the IRAF {\tt
gemini} package.

\begin{figure}
\centerline{\psfig{file=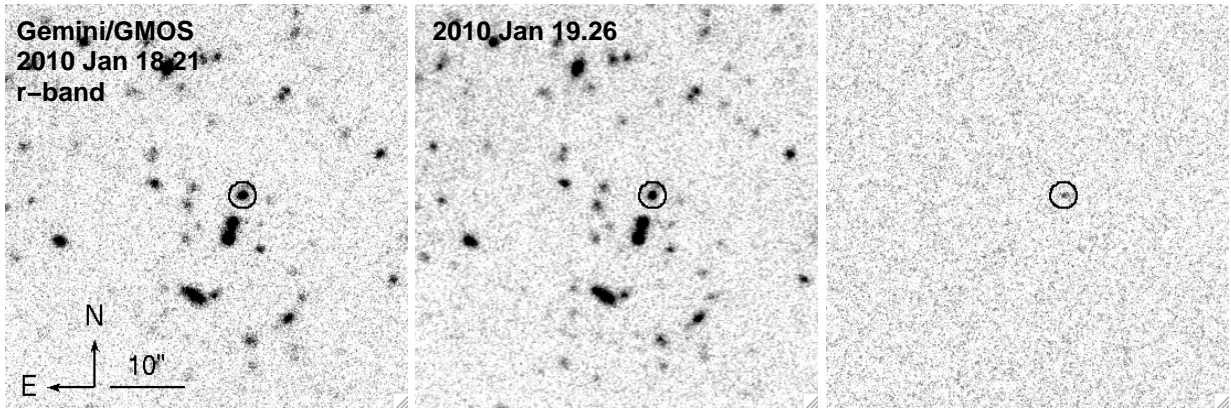,width=6.4in}}
\caption{Gemini/GMOS $r$-band images obtained starting 7.9 hr (left)
and 33.2 hr (center) after the burst.  Digital subtraction of the two
images reveals a fading source in the residual image (right), which we
identify as the afterglow.  The afterglow position is denoted by the
black circle.
\label{fig:sub}} 
\end{figure}

Digital image subtraction of the two GMOS-N observations using the
ISIS software package \citep{ala00} reveals the presence of a fading
source, which we identify as the optical afterglow of \grb; see
Figure~\ref{fig:sub}.  Assuming that the afterglow contribution is
negligible in the second GMOS-N observation we calculate a magnitude
of $r_{\rm AB}=25.46\pm 0.20$ mag ($0.24\pm 0.05$ $\mu$Jy) at a median
time of 8.3 hr after the burst.  A comparison of our IMACS observation
with the second epoch of GMOS-N imaging yields an afterglow magnitude
of $r_{\rm AB}>23.93$ mag ($3\sigma$) at a median time of 4.1 hr after
the burst.

Astrometry relative to the USNO-B catalog provides an absolute
position for the optical afterglow of $\alpha=$\ra{00}{45}{04.660},
$\delta=$\dec{-01}{35}{41.89} (J2000), with an uncertainty of $0.26''$
in each coordinate.  This position is $1.5''$ away from the center of
the XRT error circle, which has an uncertainty of $2.4''$.  We
additionally measure the relative position of the afterglow and host
galaxy (using the second Gemini epoch) and find an offset of
$\delta{\rm RA}=60$ mas and $\delta{\rm Dec}=0$ mas.  The uncertainty
in the offset includes contributions from the astrometric tie of the
two Gemini observations ($\sigma_{\rm GB\rightarrow GB}=9$ mas), the
positional accuracy of the afterglow residual ($\sigma_{\rm
\theta,GRB}=10$ mas), and uncertainty in the centroid of the host
galaxy ($\sigma_{\rm\theta,gal}=20$), which is itself dominated by
systematic uncertainty rather than just the signal-to-noise ratio.
Thus, the total angular offset between the afterglow and host center
is $60\pm 40$ mas.

\subsection{Host Galaxy Observations}
\label{sec:hostobs}

\begin{figure}
\centerline{\psfig{file=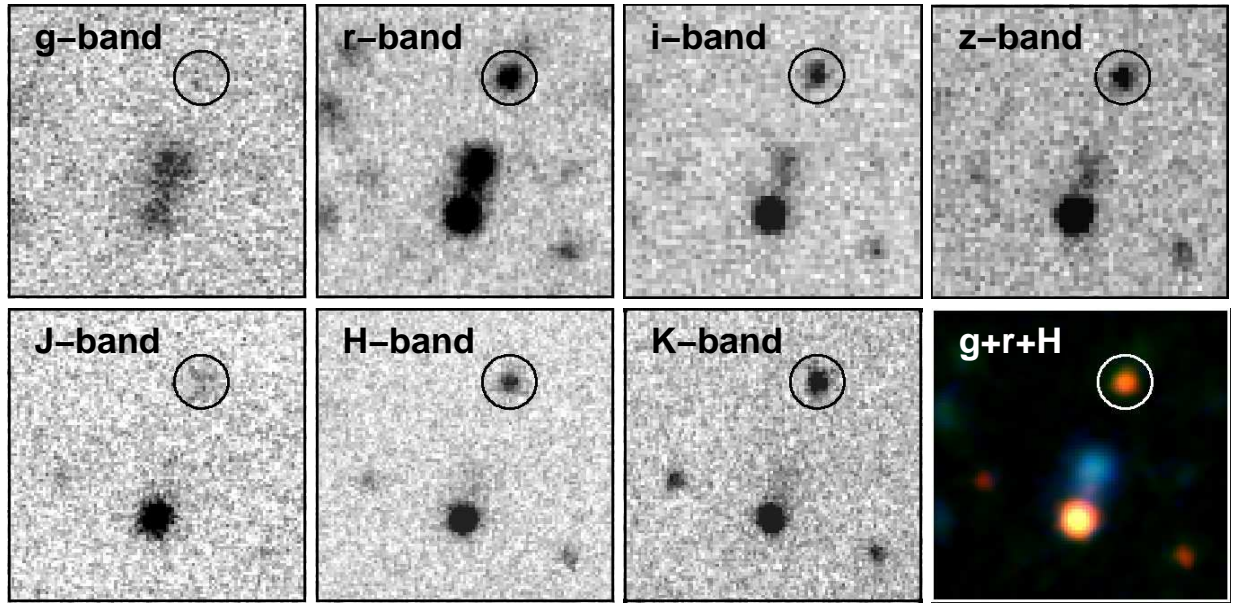,width=6.4in}}
\caption{Gemini and Magellan optical and near-IR images of the host
galaxy of \grb\ obtained with GMOS, NIRI and IMACS
(Table~\ref{tab:obs}).  Each panel is $0.2'$ on a side with an
orientation of north up and east to the left.  Also shown is a $grH$
color composite highlighting the red color of the host galaxy.
\label{fig:host}} 
\end{figure}

\begin{figure}
\centerline{\psfig{file=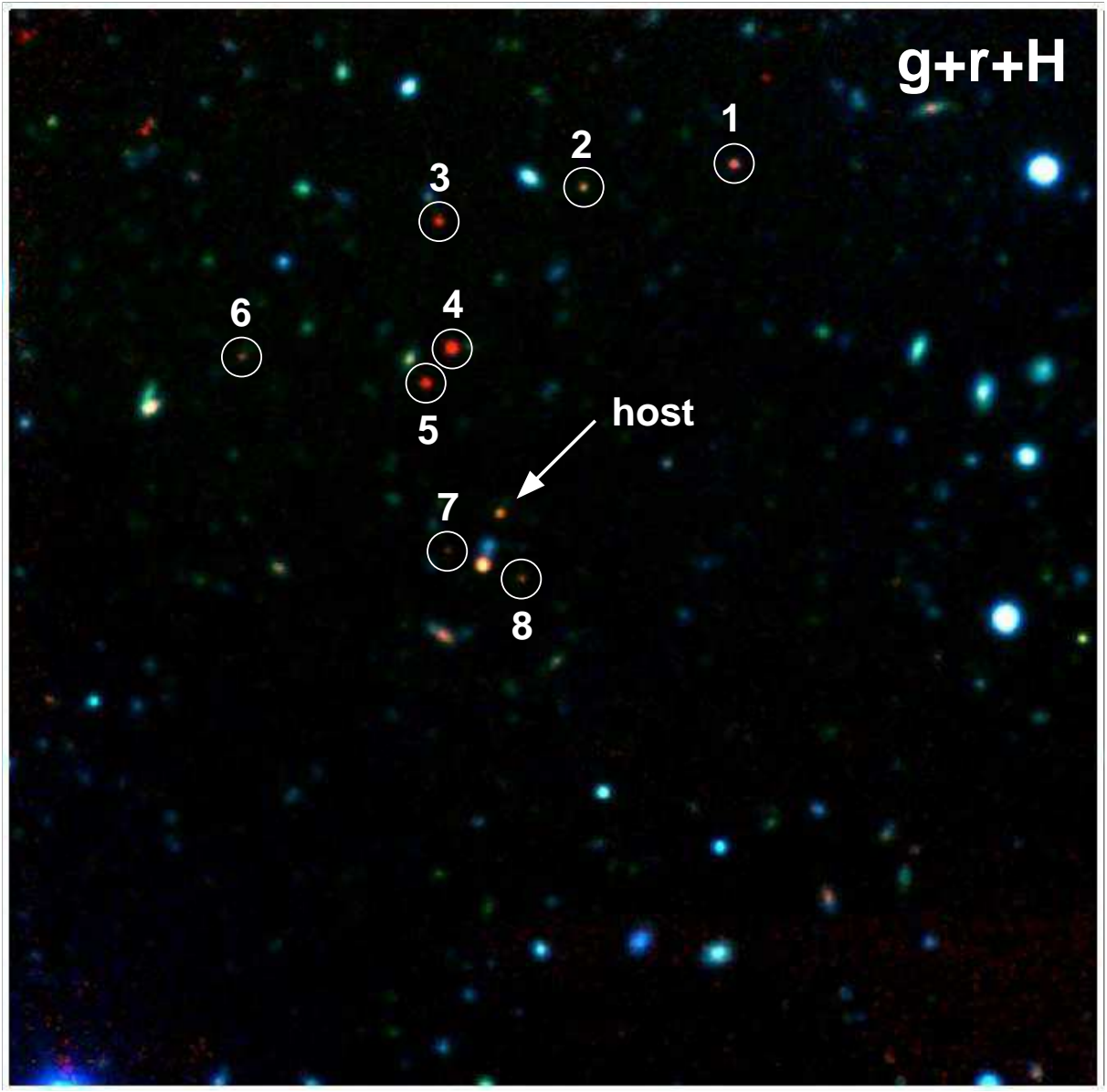,width=6.4in}}
\caption{Gemini optical and near-IR $grH$ color composite image of a
$2'\times 2'$ field around \grb.  The environment around the host
contains several red galaxies, whose colors are plotted in
Figure~\ref{fig:color} (see \S\ref{sec:hostfield}).
\label{fig:field}} 
\end{figure}

Subsequent to the discovery of the afterglow we obtained follow-up
observations of the host galaxy in the $g$-band with GMOS-N and in the
$JHK$ bands with the Near Infra-Red Imager and Spectrometer (NIRI) on
the Gemini-North 8-m telescope.  We also obtained IMACS observations
in the $riz$ bands (Table~\ref{tab:obs}).  Photometry of the host was
extracted in a $1.6''$ radius aperture, and is summarized in
Table~\ref{tab:obs}.  We note that the errors are dominated by
uncertainty in the zeropoint in the $H$ and $K$ bands.  The host
appears to be mildly resolved with a FWHM of $0.6''$ in the $K$-band
(${\rm PSF}=0.5''$). Images of the host in the $grizJHK$ filters, and
a combined color image are shown in Figure~\ref{fig:host}.  A combined
color image of the $2'$ field covered by NIRI is shown in
Figure~\ref{fig:field}.

We obtained spectroscopic observations of the host on 2010 January
19.22 using GMOS-N at a mean airmass of 1.4. A dithered pair of 1500 s
exposures were obtained with the R400 grating covering $3900-8130$
\AA\ at a spectral resolution of about 7 \AA. We obtained a second,
deeper set of observations ($4\times 1460$ s) on 2010 November 2.12
with GMOS-S on the Gemini-South 8-m telescope at a mean airmass of
1.15 in the nod-and-shuffle mode.  These observations were also
obtained with the R400 grating covering $5400-9650$ \AA\ with the
OG515 order-blocking filter. The data were reduced using standard
tasks in IRAF and wavelength calibration was performed using CuAr arc
lamps. Archival observations of the smooth-spectrum standard star EG
131 \citep{bes99} and custom IDL programs were used to apply a flux
calibration and remove the telluric absorption bands. In order to
maximize the signal-to-noise ratio, we use a weighted co-addition of
the two epochs in our subsequent analysis. The resulting spectrum is shown in
Figure~\ref{fig:hostspec}.

We detect continuum emission from the host beyond $\approx 5000$ \AA,
with a notable increase in the flux redward of $7650$ \AA.  No obvious
emission features are detected.  Interpreting the break as the
Balmer/4000\AA\ break we find an estimated redshift of $z\approx 0.9$.

\section{Host Galaxy Redshift and Properties} 
\label{sec:host}

\subsection{Spectroscopy} 
\label{sec:hostspec}

\begin{figure} 
\plottwo{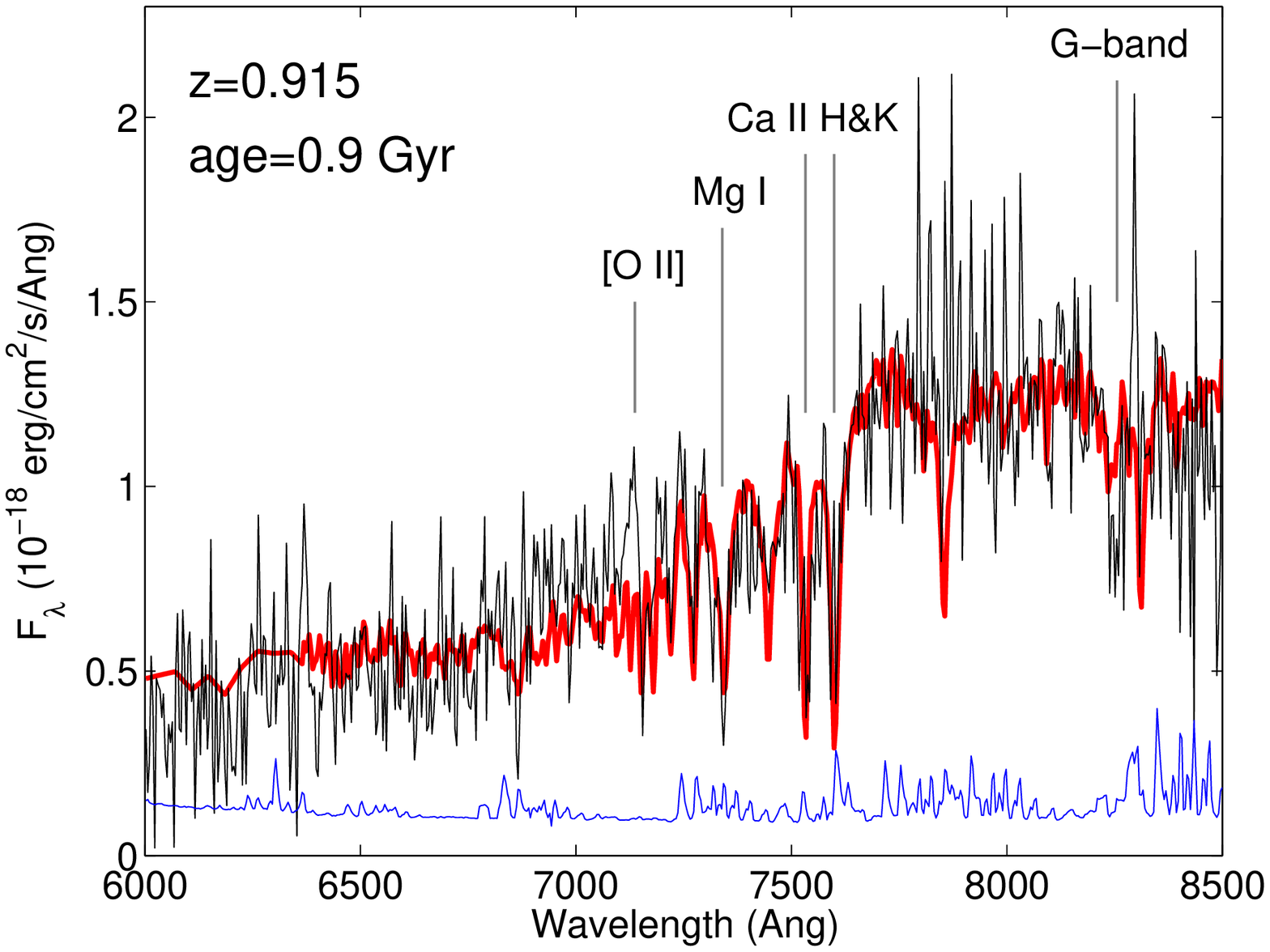}{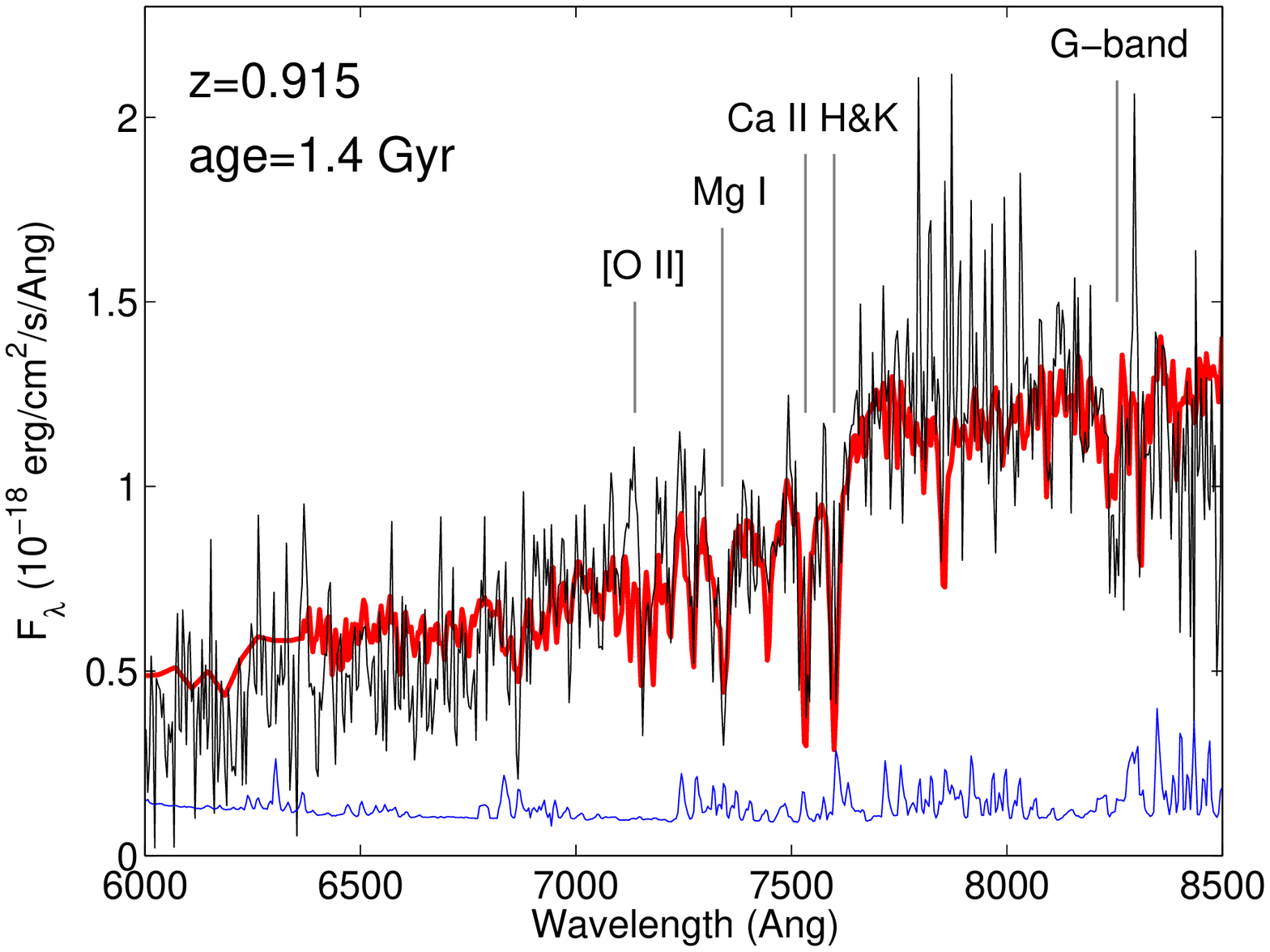}
\plottwo{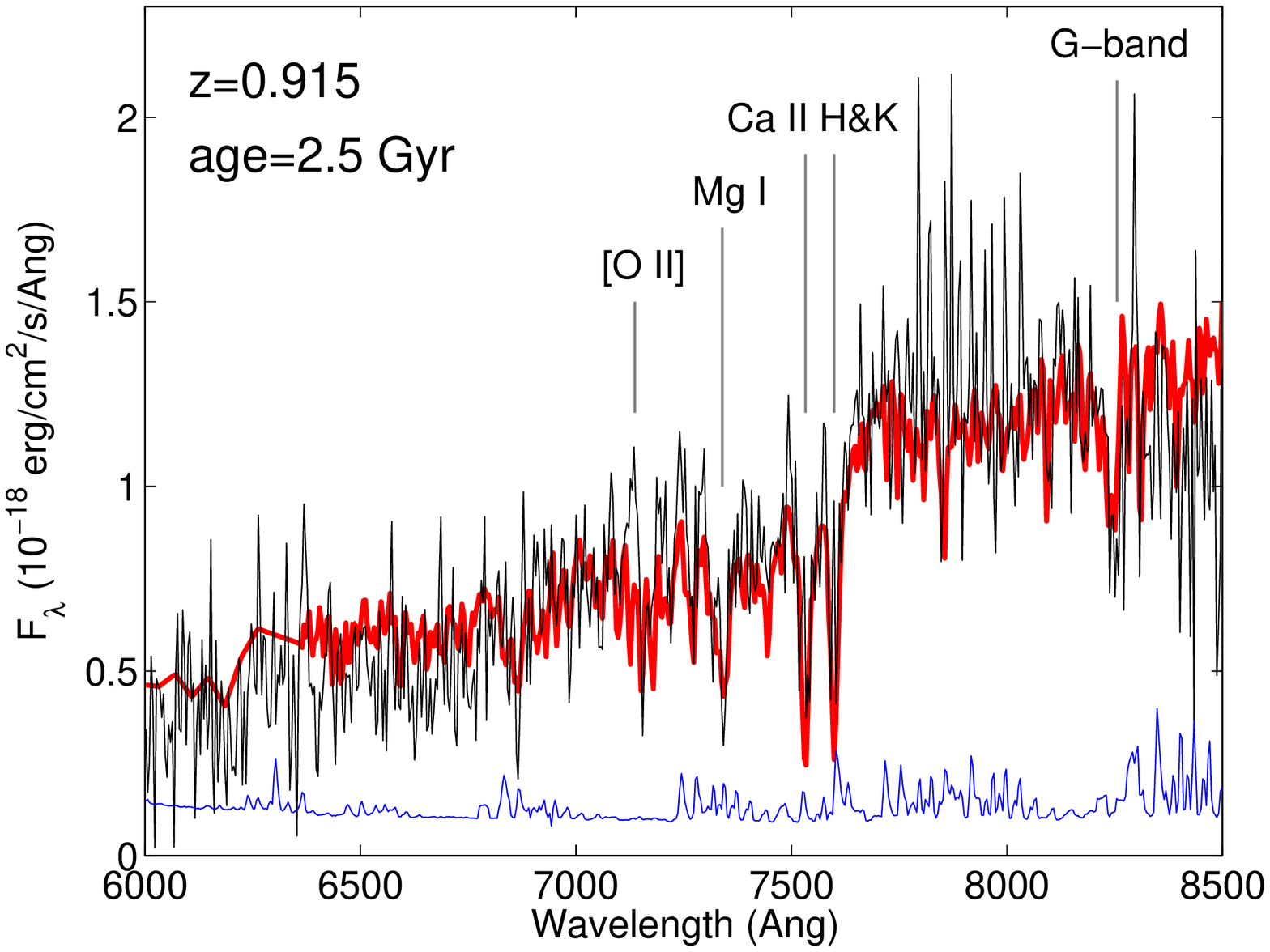}{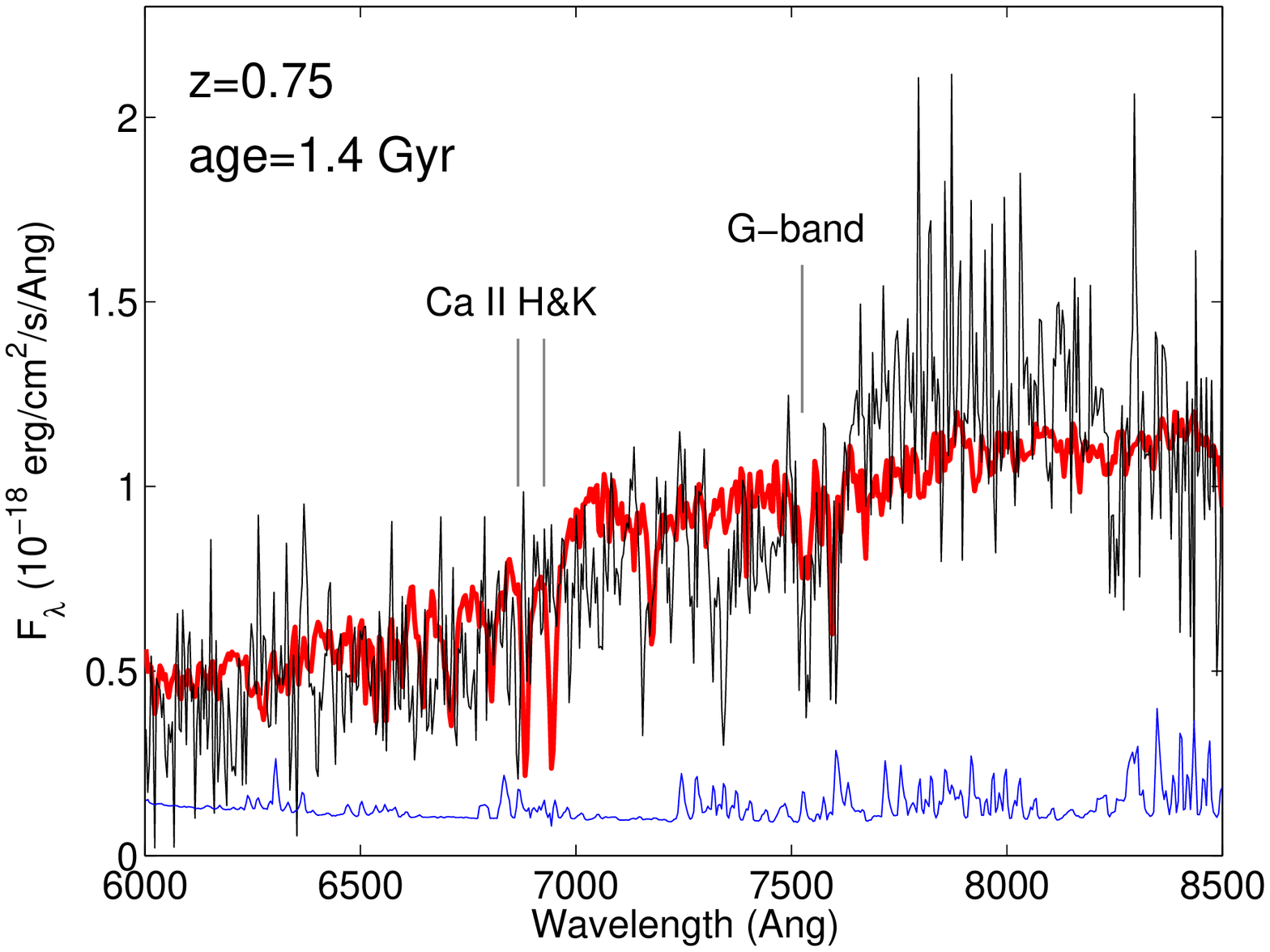}
\caption{Gemini spectrum of the host galaxy of \grb\ binned with
3-pixel boxcar (black: data; blue: error spectrum).  Also shown are
SSP templates (red; \citealt{bc03}) with stellar population ages of
0.9 Gyr (top left), 1.4 Gyr (top right), and 2.5 Gyr (bottom left) at
the best-fit spectroscopic redshift of $z=0.915$.  We also show a 1.4
Gyr SSP template at the preferred photometric redshift of $z=0.75$
(bottom right).  Fits are performed on the unbinned data.  The latter
fit provides a poorer match to both the sharpness of the break and the
main spectral features.  Absorption line locations corresponding to
\ion{Ca}{2} H\&K, \ion{Mg}{1}$\lambda 3830$, and G-band$\lambda 4300$
are indicated.  Also shown is the expected location of the
[\ion{O}{2}]$\lambda 3727$ emission doublet.
\label{fig:hostspec}} 
\end{figure}

\begin{figure}
\centerline{\psfig{file=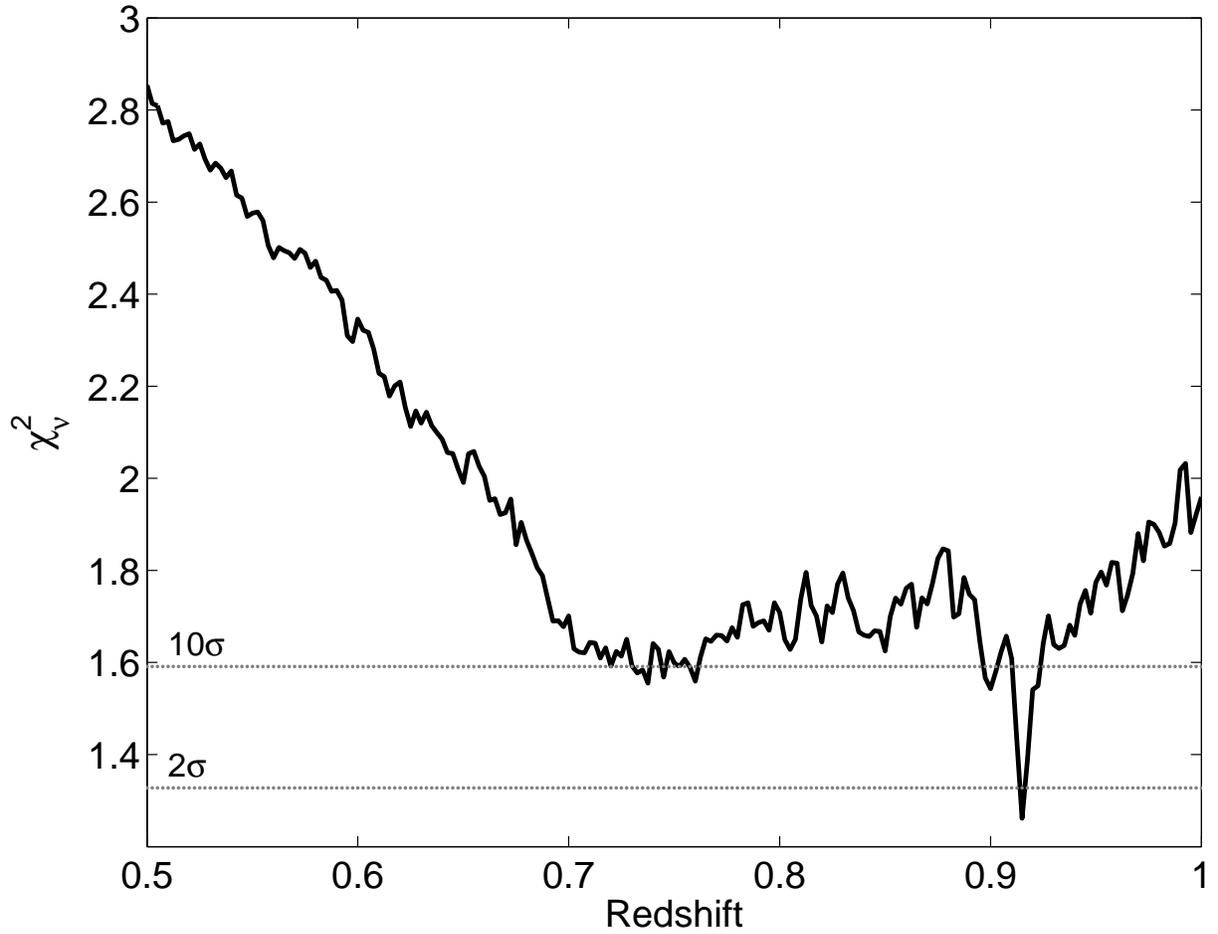,width=6.4in}}
\caption{$\chi_{\nu}^2$ as a function of redshift for the
spectroscopic fit to a 1.4 Gyr SSP template (performed on the unbinned
data, described in \S\ref{sec:hostspec}).  The $2\sigma$ and
$10\sigma$ levels are labeled.  We find a sharp minimum at $z=0.915$
with $\chi^2_{\nu}=1.26$ and a broad minimum at $z\sim 0.75$, which is
only consistent with the data at the $10\sigma$ level.
\label{fig:chisq1}} 
\end{figure}

To quantitatively assess the host galaxy's redshift, we fit a weighted
co-addition of the GMOS spectra described in \S\ref{sec:hostobs} over
the wavelength range $6000-8500$ \AA\ with spectral evolution models
of simple stellar populations (SSPs) provided as part of the GALAXEV
library \citep{bc03}; at wavelengths outside this range, the
signal-to-noise ratio is too low to contribute significantly to the
fit.  We use $\chi^{2}$ minimization with redshift as the free
parameter and the best-fit flux normalization determined by the
equation,
\begin{equation}
F_{\rm 0,bf}=\frac {\sum_{i=1}^n \frac{F_{\lambda,model,i} \times
F_{\lambda,gal,i}}{\sigma^2_{\lambda,gal,i}}}{\sum_{i=1}^n
\frac{F_{\lambda,model,i}^2}{\sigma^2_{\lambda,gal,i}}},
\end{equation}
where $F_{\lambda,model,i}$ are the model fluxes, and
$F_{\lambda,gal,i}$ and $\sigma_{\lambda,gal,i}$ are the observed
galaxy fluxes and uncertainties, respectively.  The fit is performed
on the unbinned data.  The resulting best-fit redshift is $z=0.915$
($\chi^{2}_{\nu}=1.26$ for $1841$ degrees of freedom at 1.4 Gyr) for
SSPs with an age of $0.9-2.5$ Gyr; see Figure~\ref{fig:hostspec}.
Significantly poorer fits are found with SSPs outside of this age
range, or with late-type templates.  At this redshift, there is a
clear match between absorption features in the spectrum and the
expected dominant lines (i.e., \ion{Ca}{2} H\&K, \ion{Mg}{1}, and
G-band).  The distribution of $\chi^{2}_\nu$ as a function of redshift
is shown in Figure~\ref{fig:chisq1} revealing a secondary broad
minimum at $z\approx 0.75$.  This solution provides a much poorer fit
to the data with $\chi^{2}_{\nu}=1.60$, corresponding to about
$10\sigma$ away from the best fit.  The main reason for the poor fit
is that it misses the key spectral absorption features and the clear
break at 7650 \AA\ (Figure~\ref{fig:hostspec}).

At the best-fit spectroscopic redshift of $z=0.915$ we find a marginal
feature corresponding to [\ion{O}{2}]$\lambda 3727$ with a flux of
$F_{\rm [OII]}\approx 3\times 10^{-18}$ erg cm$^{-2}$ s$^{-1}$, or
$L_{\rm [OII]}\approx 7\times 10^{39}$ erg s$^{-1}$.  Given the
marginal detection we use this luminosity to derive an upper limit on
the star formation rate, with ${\rm SFR}=(1.4\pm 0.4)\times
10^{-41}\,L_{\rm [OII]}\lesssim 0.1$ M$_\odot$ yr$^{-1}$
\citep{ken98}.  This is smaller than the star formation rates inferred
for the star forming hosts of short GRBs \citep{ber09}.

\subsection{Broad-band Photometry}
\label{sec:hostphot}

\begin{figure}
\includegraphics[angle=0,width=3.6in]{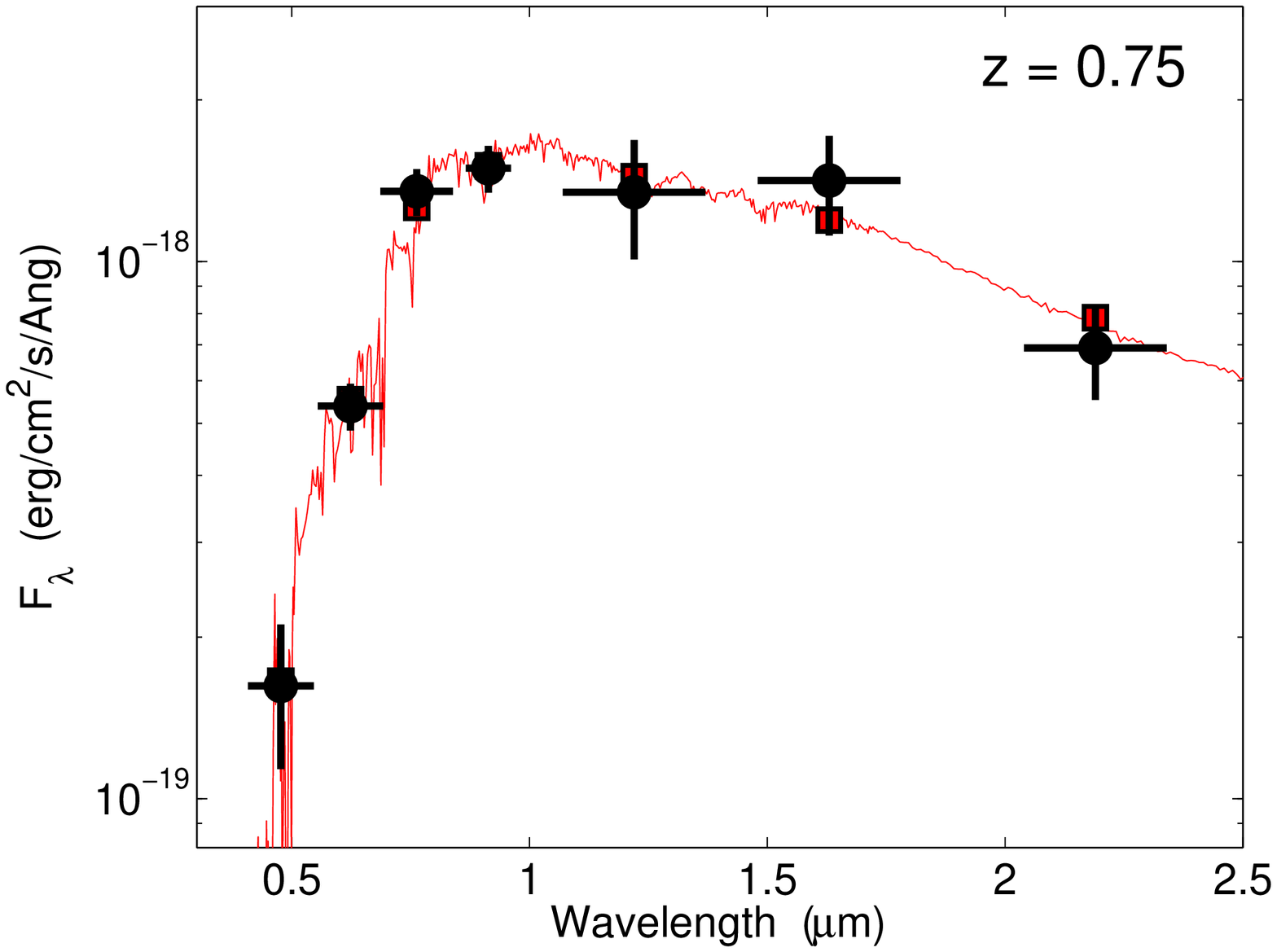}
\includegraphics[angle=0,width=3.6in]{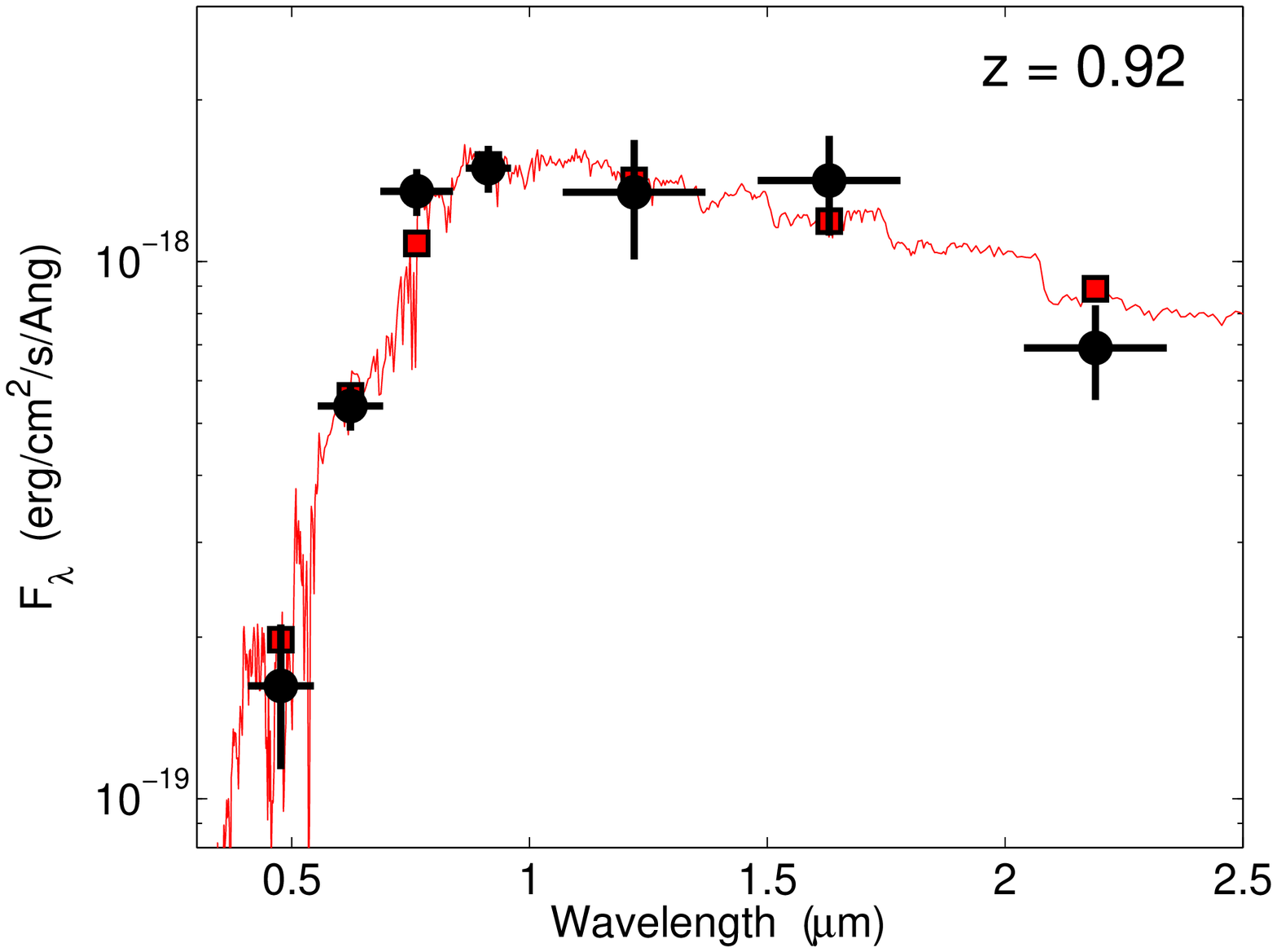}
\caption{Optical and near-IR spectral energy distribution of the host
galaxy of \grb\ (black circles).  Each SED is fit with a \citet{mar05}
single stellar population model (red line) through a maximum
likelihood fit of the synthesized photometry (red squares,
\S\ref{sec:hostphot}).  We show the fits at the photometric best-fit
$z=0.75$ (left), and also at the preferred spectroscopic redshift
$z=0.915$ (right).
\label{fig:sed}} 
\end{figure}

\begin{figure}
\centerline{\psfig{file=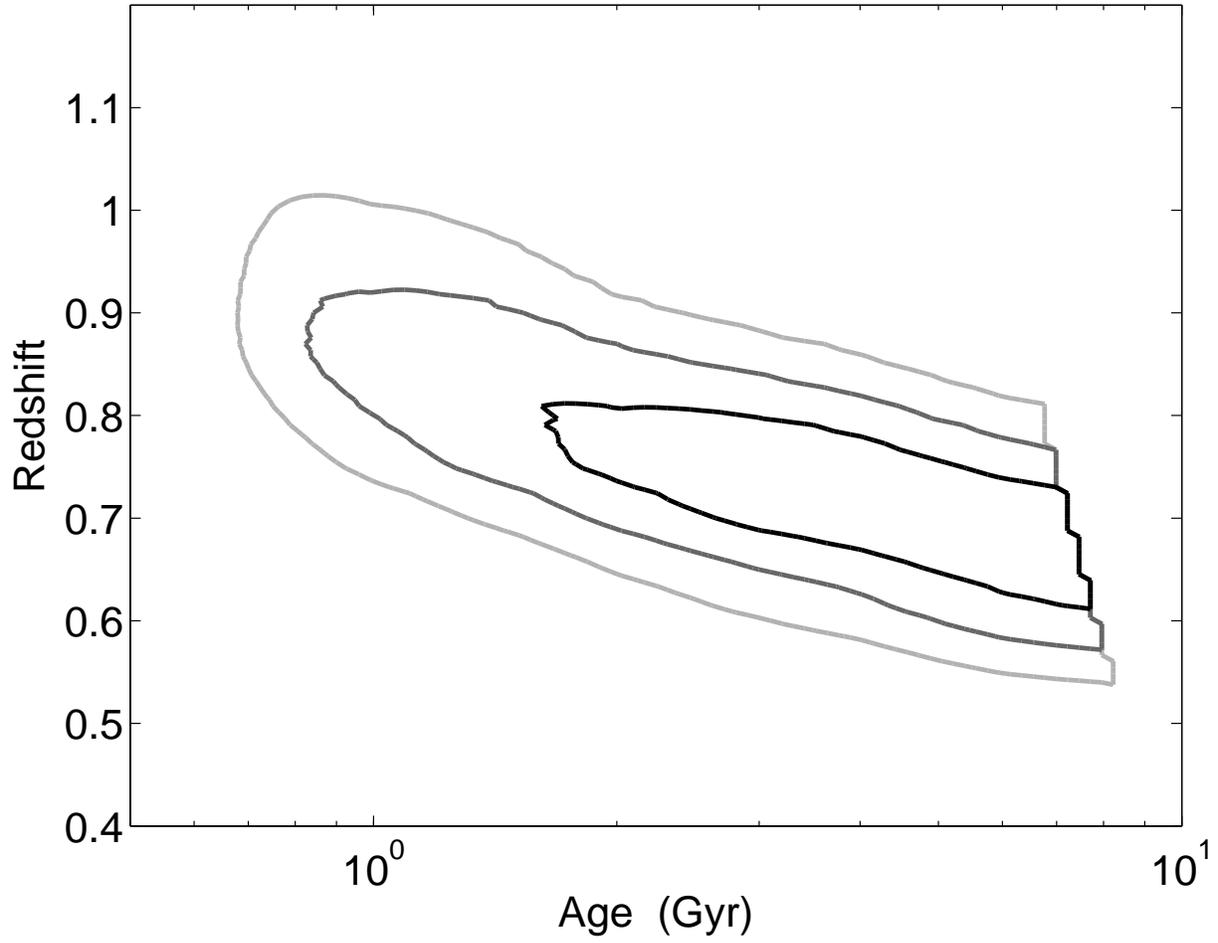,width=6.4in}}
\caption{$\chi_{\nu}^2$ contours for redshift and stellar population
age using the \citet{mar05} single stellar population model
($1,2,3\sigma$ in order of decreasing darkness).  The jagged edge at
large ages is due to truncation at the appropriate age of the universe
as a function of redshift.  The $2\sigma$ contour leads to a best-fit
redshift of $z=0.57-0.92$.
\label{fig:chisq2}} 
\end{figure}

To extract additional information about the host galaxy we fit the
broad-band photometry with \citet{mar05} evolutionary stellar
population synthesis models.  We use the subset of models described in
\citet{lb10}, with a Salpeter initial mass function, solar
metallicity, and a red horizontal branch morphology, leaving redshift
as a free parameter; see Figure~\ref{fig:sed}.  We find the best-fit
solution to be $z=0.75$ ($\chi^2_\nu=0.3$ for 5 degrees of freedom)
with a $2\sigma$ range of $z\approx 0.57-0.92$
(Figure~\ref{fig:chisq2}), consistent with our spectroscopic redshift
determination.  Since $z=0.75$ is ruled out at high confidence from
the spectroscopic fit, we adopt $z=0.915$ as the redshift of the host.
At this redshift, the inferred stellar mass is $2.6\times 10^{10}$
M$_\odot$ and the stellar population age is about $1.1$ Gyr, in good
agreement with the spectroscopic results.  The absolute B-band
magnitude is M$_{\rm B}\simeq -20.3$ mag, corresponding to L$_{\rm
B}\simeq 0.5$L$_{\rm *}$ in comparison to the DEEP2 luminosity
function at $z=0.9$ \citep{wfk+06}. We generally find poorer fits for
models with $0.5$ and $2$ Z$_{\odot}$.

Combining the upper limit on the star formation rate with the inferred
stellar mass, the resulting limit on the specific star formation rate
is ${\rm SSFR}\equiv {\rm SFR}/M_*\lesssim 0.004$ Gyr$^{-1}$.  This
confirms that the host galaxy is quiescent, since the characteristic
growth timescale, ${\rm SSFR}^{-1}\approx 260$ Gyr, is significantly
larger than the Hubble time.

\subsection{Large-Scale Environment}
\label{sec:hostfield}

\begin{figure}
\centerline{\psfig{file=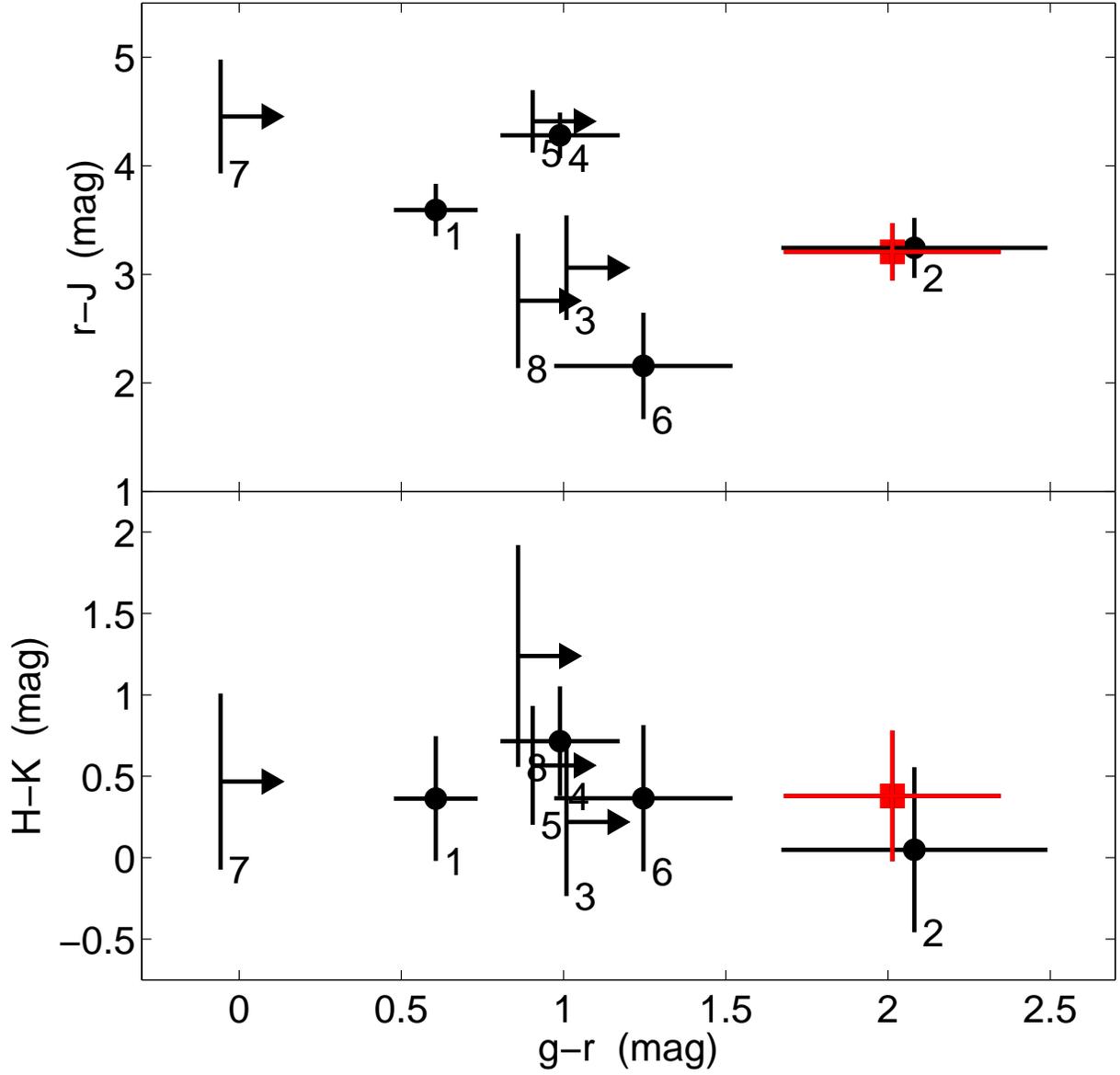,width=6.4in}}
\caption{Color-color plots for red galaxies in a $2'\times 2'$ field
around \grb\ (Figure~\ref{fig:field}).  The host galaxy is marked by a
red square.  We identify an additional galaxy with similar colors
(\#2), and two potential galaxies with similar colors (although only
lower limits in $g-r$; \#3 and \#8).
\label{fig:color}} 
\end{figure}

As shown in Figure~\ref{fig:field}, the field around \grb\ contains
several red galaxies in addition to the host galaxy itself.  We
investigate whether the host and these galaxies are part of a
large-scale structure similar to some previous short GRB environments
\citep{bpp+06,bsm+07}, using color-color plots (Figure~\ref{fig:color}).
We find that only one galaxy has similar colors to the host, and is
therefore potentially located at the same redshift.  Two additional
galaxies have similar $r-J$ and $H-K$ colors to the host, with limits
on their $g-r$ colors (due to $g$-band non-detections) that are
consistent with the host of \grb.  These galaxies may be less luminous
members of the same large-scale structure.  Additional spectroscopic
observations are required to assess whether the host is part of a
$z\approx 0.9$ galaxy group. However, we note that even if it is part
of a real group, it is not a rich group or cluster.

\section{GRB and Afterglow Properties} 
\label{sec:ag}

At the best-fit redshift of $z=0.915$ the isotropic-equivalent
$\gamma$-ray energy of \grb\ is $E_{\rm\gamma,iso}=9.2\times 10^{50}$
erg ($16-2000$ keV in the rest-frame).  This is similar to the values
inferred for previous short GRBs at a similar redshift
\citep{bfp+07,ber07,ber10a}.

To extract additional information about the burst we use the measured
brightness of the optical afterglow in conjunction with the limit on
the X-ray flux (\S\ref{sec:obs}).  The inferred optical to X-ray
spectral index is $\beta_{\rm OX}\lesssim -0.75$, in agreement with
the range of $\langle\beta_{\rm OX}\rangle=-0.72\pm 0.17$ measured for
short GRBs with X-ray and optical afterglow detections
\citep{nfp09,ber10a}.  Assuming the standard synchrotron emission
spectrum from a relativistic blast wave \citep{spn98}, this value of
$\beta_{\rm OX}$ indicates $p\gtrsim 2.5$ if $\nu_c>\nu_{\rm X}$, or
$p\gtrsim 1.5$ if $\nu_c<\nu_{\rm O}$; here $\nu_c$ is the synchrotron
cooling frequency and $p$ is the power law index of the electron
energy distribution, $N(\gamma)\propto\gamma^{-p}$.  Since these
values are not atypical, we cannot robustly locate $\nu_c$ based on
the optical to X-ray flux ratio.

However, we can still constrain the circumburst density ($n$) by
making the reasonable assumption that the cooling frequency is located
above the optical band, while the synchrotron peak frequency, $\nu_m$,
is located below the optical band since the optical afterglow is
fading at discovery.  Using the observed optical afterglow brightness
and assuming a constant density medium we find \citep{gs02}:
\begin{equation}
n\approx 7.3\times 10^{-12}\epsilon_e^{-3}\epsilon_B^{-1.75}
E_{52}^{-2.75}\,\,\,{\rm cm}^{-3},
\label{eqn1}
\end{equation}
where $\epsilon_e$ and $\epsilon_B$ are the fractions of energy in the
radiating electrons and magnetic field, respectively, and $E_{52}$ is
the energy in units of $10^{52}$ erg.  Assuming that $E\approx
E_{\rm\gamma,iso}$ and using $\epsilon_e,\epsilon_B\lesssim 1/3$ we
infer a lower limit on the density of $n\gtrsim 10^{-6}$
cm$^{-3}$, which is similar to IGM densities. For more typical values
of $\epsilon_e\approx\epsilon_B \approx 0.1$, we find $n\approx
3\times 10^{-4}$ cm$^{-3}$.  Since generally
$\epsilon_e,\epsilon_B\gtrsim 0.01$ \citep{pk02,yhs+03}, a likely
upper bound on the density is $n\lesssim 20 $ cm$^{-3}$.

\section{Discussion and Conclusions}
\label{sec:discussion}

In the sample of 14 short GRBs with optical afterglows and coincident
hosts, \grb\ is only the second event unambiguously associated with an
early-type galaxy (the other being GRB\,050724; \citealt{bpc+05}).
Additional cases of early-type hosts have been proposed.  In
particular, GRB\,050509b is likely associated with an early-type
cluster galaxy but this is based on only an X-ray position
\citep{bpp+06}.  Two additional bursts (070809 and 090515) lack
coincident hosts despite optical afterglow detections, but in both
cases the galaxies with the lowest probability of chance coincidence
are early-type galaxies \citep{ber10a}.  Even if we accept these
additional early-type host associations as genuine, the host of \grb\
is located at a significantly higher redshift than the previous
events, with $z\approx 0.23-0.47$.  \grb\ also has the highest
isotropic-equivalent gamma-ray energy of these events by a factor of a
few, with $E_{\rm\gamma,iso}=2.1 \times 10^{50}$ erg ($15-150$
keV). These results suggest that some of the optically-faint host
galaxies identified to date (e.g., \citealt{bfp+07}) may be bright
near-IR sources due to a dominant old population.  It also indicates
that the presence of short GRBs in early-type galaxies does not
necessarily point to progenitor ages of $\sim 10$ Gyr.  Instead, the
typical ages of short GRB progenitors in early-type hosts appear to be
$\sim 1-4$ Gyr \citep{lb10}, which may lead to early-type hosts even
at $z\approx 3$.

\begin{figure}
\centerline{\psfig{file=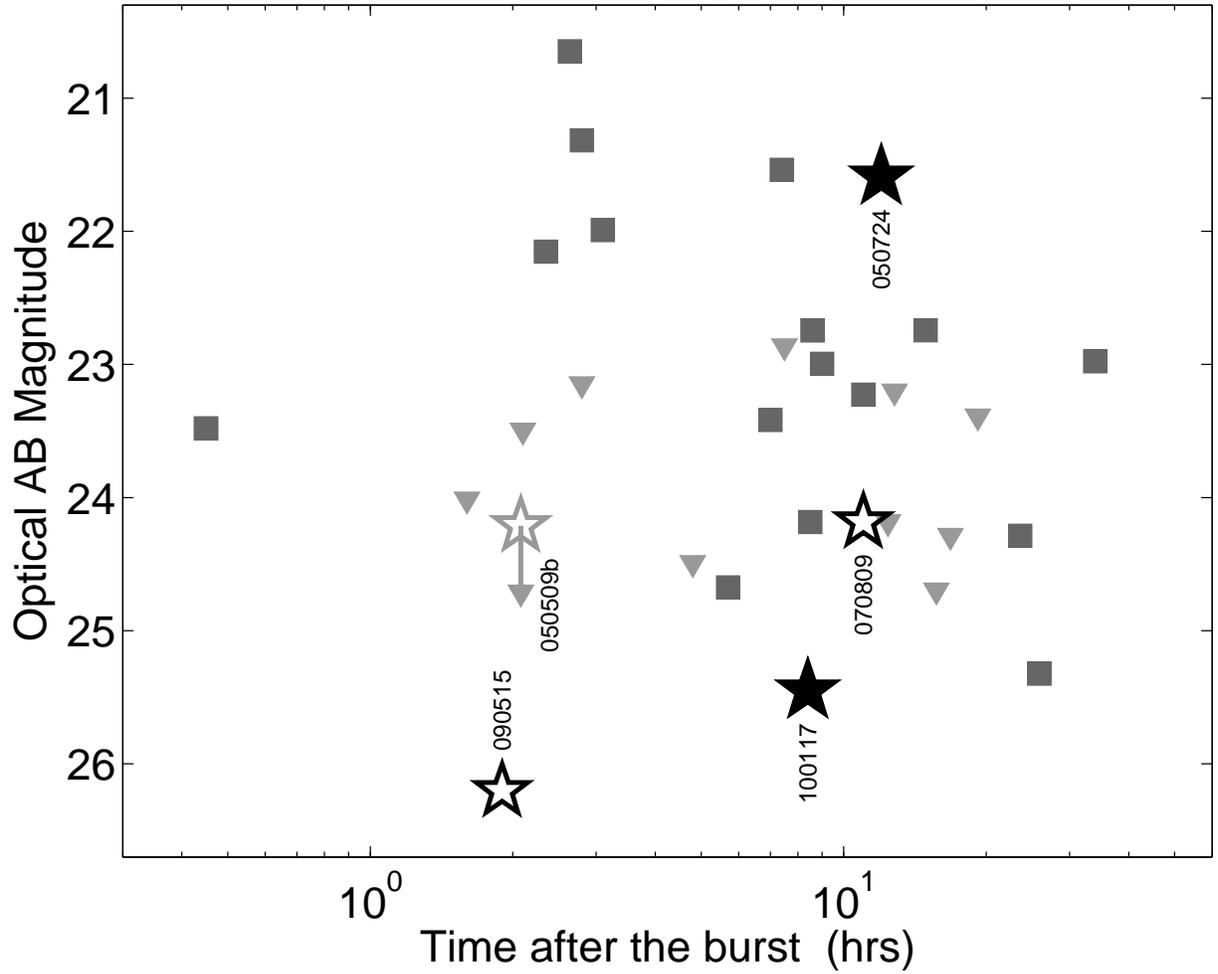,width=6.4in,angle=0}}
\caption{Short GRB optical afterglow brightness at the time of
discovery for bursts with detected optical afterglows (squares) and
upper limits (arrows).  The two short GRBs with secure early-type
hosts are denoted by solid stars, while bursts with putative
early-type hosts are marked by open stars (\citealt{ber10a} and
references therein).  Short GRBs with early-type hosts may have weaker
optical afterglows on average, possibly related to lower circumburst
densities.
\label{fig:optag}} 
\end{figure}

At the inferred redshift of $z=0.915$ the projected physical offset of
\grb\ is only $470\pm 310$ pc.  Our previous analysis of short GRB
offsets revealed a median projected offset of about 5 kpc
\citep{fbf10}.  In this context, \grb\ has the smallest offset
measured to date.  We note that the only other burst with a secure
early-type host (GRB\,050724) also has a small offset of about $2.7$
kpc.  Given the age of the stellar population of $\sim 1$ Gyr in both
cases (see also \citealt{lb10}), these small offsets indicate that
\grb\ and GRB\,050724 did not originate from progenitors with a
substantial kick (unless the kick direction in both cases is nearly
aligned with our line of sight).  Given the lack of any recent star
formation activity, we can also rule out the possibility of a highly
kicked progenitor system with a short merger time.  On the other hand,
the proposed associations of short GRBs 050509b, 070809, and 090515
with early-type hosts at offsets of tens of kpc \citep{bpp+06,ber10a}
indicates that some progenitors may experience large kicks.  The cases
of GRBs 070809 and 090515 is particularly intriguing since both had
optical afterglows of comparable brightness to \grb\
(Figure~\ref{fig:optag}), suggestive of a similar circumburst density
despite a potential large difference in offsets.

Only a few short GRBs have circumburst density measurements, reflecting
a general lack of multi-wavelength afterglow detections.  GRB\,051221A
had an estimated density of $n\sim 10^{-3}$ cm$^{-3}$ \citep{sbk+06},
GRB\,050724 had $n\approx 0.01-0.1$ cm$^{-3}$ \citep{bpc+05}, and
GRB\,050709 had\footnotemark\footnotetext{Only an upper bound is
available due to the lack of a radio detection.} $n\lesssim 0.1$
cm$^{-3}$ \citep{pan06}.  For \grb\ we estimate $n\sim 10^{-4}-10$
cm$^{-3}$, continuing the trend of relatively low circumburst
densities for short GRBs.  This is particularly striking in comparison
to the circumburst densities inferred for long GRBs, with a median of
$\langle n\rangle\approx 1-10$ cm$^{-3}$ (e.g., \citealt{sbk+06}).

Our discovery of the afterglow and $z=0.915$ early-type host of \grb\
continues to support the conclusion that short GRBs exist at $z\sim 1$
and beyond \citep{bfp+07}.  However, unlike all previous short GRB
hosts at these redshifts \citep{bfp+07,gfl+09,aap+09,lbb+10}, the host
of \grb\ exhibits no evidence for star formation activity and is
instead dominated by a $\sim 1$ Gyr old stellar population.  With its
faint optical afterglow it is possible that previous such events have
been missed due to shallow optical afterglows searches, thereby
potentially biasing the known host population against $z\gtrsim 1$
early-type hosts.

\acknowledgements We thank Rik Williams and Daniel Kelson for
obtaining rapid observations of \grb\ with IMACS. This paper
includes data gathered with the 6.5 meter Magellan Telescopes located
at Las Campanas Observatory, Chile.  Observations were also obtained
at the Gemini Observatory, which is operated by the Association of
Universities for Research in Astronomy, Inc., under a cooperative
agreement with the NSF on behalf of the Gemini partnership: the
National Science Foundation (United States), the Particle Physics and
Astronomy Research Council (United Kingdom), the National Research
Council (Canada), CONICYT (Chile), the Australian Research Council
(Australia), CNPq (Brazil) and CONICET (Argentina). This work also
made use of data supplied by the UK Swift Science Data Center at the
University of Leicester.  This work was partially supported by Swift
AO5 grant \#5080010 and AO6 grant \#6090612.

\clearpage
\begin{deluxetable}{lcccccccccc}
\tabletypesize{\footnotesize}
\tablecolumns{10}
\tabcolsep0.05in\footnotesize
\tablewidth{0pc}
\tablecaption{Log of Optical and Near-IR and Observations of \grb\
\label{tab:obs}}
\tablehead {
\colhead {Date}                &
\colhead {$\Delta t$}          &
\colhead {Telescope}           &
\colhead {Instrument}          &
\colhead {Filter}              &
\colhead {Exposures}           &
\colhead {$\theta_{\rm FWHM}$} &
\colhead {Afterglow$^a$}       &
\colhead {$F_\nu^a$}           &
\colhead {Host$^a$}            &
\colhead {$A_{\lambda}^b$}     \\
\colhead {(UT)}                &
\colhead {(d)}                 &
\colhead {}                    &
\colhead {}                    &
\colhead {}                    &
\colhead {(s)}                 &
\colhead {($''$)}              &
\colhead {(AB mag)}            &
\colhead {($\mu$Jy)}           &
\colhead {(AB mag)}            &
\colhead {(mag)}                                   
}
\startdata
2010 Jan 18.040 & 0.162 & Magellan     & IMACS & $R$ & $4\times 300$  & $1.25$ & $>23.93$         & $< 0.97$       & \nod            & $0.064$ \\
2010 Jan 18.207 & 0.329 & Gemini-North & GMOS  & $r$ & $15\times 180$ & $0.80$ & $25.46\pm 0.20$  & $0.24\pm 0.05$ & \nod            & $0.07$  \\
2010 Jan 19.262 & 1.383 & Gemini-North & GMOS  & $r$ & $15\times 180$ & $0.76$ & \nod             & $0^{c}$        & $24.30\pm 0.10$ & $0.07$  \\
2010 Feb 2.208  & 15.3  & Gemini-North & NIRI  & $K$ & $19\times 60$  & $0.51$ &                  &                & $21.24\pm 0.20$ & $0.01$  \\
2010 Feb 2.229  & 15.4  & Gemini-North & NIRI  & $H$ & $19\times 60$  & $0.56$ &                  &                & $21.26\pm 0.21$ & $0.01$  \\
2010 Feb 4.208  & 17.3  & Gemini-North & NIRI  & $J$ & $14\times 60$  & $0.80$ &                  &                & $21.87\pm 0.25$ & $0.02$  \\
2010 Feb 4.229  & 17.4  & Gemini-North & GMOS  & $g$ & $7\times 240$  & $1.15$ &                  &                & $26.17\pm 0.30$ & $0.1$   \\
2010 Nov 14.042 & 300.2  & Magellan   & IMACS & $z$ & $5\times 180$   & $0.70$ &                  & 		   & $22.33\pm 0.10$ & $0.035$ \\
2010 Nov 14.057 & 300.2  & Magellan   & IMACS & $i$ & $3\times 240$   & $0.63$ &		  &		   & $22.85\pm 0.10$ & $0.045$ \\
2010 Nov 14.083 & 300.2  & Magellan   & IMACS & $r$ & $8\times 360$   & $0.61$ &		  &		   & $24.33\pm 0.10$ & $0.07$
\enddata
\tablecomments{$^a$ These values are corrected for Galactic extinction. \\
$^b$ Galactic extinction. \\
$^c$ We assume the afterglow contribution is negligible 1.383 d after
the burst.}
\end{deluxetable}

\clearpage

\clearpage

\clearpage

\clearpage 

\clearpage

\clearpage 

\clearpage

\clearpage

\clearpage

\end{document}